\documentstyle[12pt]{article}
\newcommand{\beq}{\begin{equation}}
\newcommand{\eeq}{\end{equation}}
\newcommand{\bea}{\begin{eqnarray}}
\newcommand{\eea}{\end{eqnarray}}

\newcommand{\tr}{{\rm tr}}
\newcommand{\V}{{\cal V}}
\newcommand{\vev}[1]{\Big\langle #1 \Big\rangle}
\input epsf
\begin{document}

\hfill \vbox{\hbox{UCLA/98/TEP/15}
             \hbox{COLO-HEP-412}
             \hbox{hep-lat/9807027}} 
\begin{center}{\Large\bf Vortices and the SU(3) 
string tension }\\[2cm] 
{\bf Tam\'as G. Kov\'acs}\footnote{Research supported by 
DOE grant DE-FG02-92ER-40672} \\
{\em Department of Physics, University of Colorado, Boulder 
CO 80309-390}\\
{\sf e-mail: kovacs@eotvos.Colorado.EDU}\\[5mm] 

and\\[5mm]

{\bf E. T. Tomboulis}\footnote{Research supported by 
NSF grant NSF-PHY 9531023}\\
{\em Department of Physics, UCLA, Los Angeles, 
CA 90095-1547}\\
{\sf e-mail: tombouli@physics.ucla.edu}
\end{center}
\vspace{1cm}

\begin{center}{\Large\bf Abstract}
\end{center}
We present simulation results comparing the $SU(3)$ heavy quark 
potential extracted from the full Wilson loop expectation 
to that extracted  from the expectation of the Wilson loop 
fluctuation solely by elements of $Z(3)$. The two potentials are found to 
coincide. This agreement is stable under multiple smoothings of 
the configurations which remove short distance fluctuations, and  
thus reflects long-distance physics. It strongly indicates that the 
asymptotic string tension arises from thick center  
vortices linking with the Wilson loop.

\vfill
\pagebreak

Recently, strong numerical evidence has been obtained 
for the vortex picture of confinement \cite{KT1}, \cite{Det}. 
In these simulations the heavy quark potential for the 
gauge group $SU(2)$ was found to be fully recovered 
solely from the $Z(2)$ part of 
the fluctuation of the Wilson loop which is caused by 
center vortices. Furthermore, an analytical 
proof was recently obtained \cite{KT2} to show that 
forbidding the linkage of thick vortices 
with a Wilson loop results into perimeter law behavior 
at weak coupling. In other words, the presence of such 
vortices linking with the loop is necessary for 
confinement at (arbitrarily) weak coupling.  

In this paper we present simulations extending the 
results in \cite{KT1} to the $SU(3)$ gauge group. 
We compute the string tension extracted from the expectation 
of the fluctuation in the value of the Wilson loop observable by 
elements of $Z(3)$. We find that it fully reproduces the asymptotic 
string tension extracted from the full Wilson loop. Furthermore, 
and most importantly, this agreement is robust under multiple 
smoothings of the configurations eliminating short distance 
fluctuations. Passing this test is indeed 
necessary if such a coincidence truly reflects long-distance physics. 
Our $SU(3)$ results then replicate 
those found to hold for $SU(2)$ \cite{KT1}. Thus, in addition to 
addressing the physically realistic case $N=3$ among the 
$SU(N)$ groups, they strongly 
suggest that a similar  state of affairs should hold for any low $N$.

To fix ideas let us recall that vortices are characterized 
by multivalued singular gauge transformations $V(x)\in SU(N)$. 
The multivaluedness ambiguity lies in the center $Z(N)$, so 
the transformation is singled-valued in $SU(N)/Z(N)$. Such a 
$V(x)$ becomes singular on  a closed surface $\V$ of codimension 
2 (i.e. a closed  loop in $d=3$, a closed 2-dimensional sheet in 
$d=4$) forming the topological obstruction 
to a single-valued choice of 
$V(x)$ throughout spacetime. Vortex 
configurations of the gauge potentials consist of a  
core region enclosing $\V$, and a pure-gauge long-range tail 
given by $V(x)$. The asymptotic pure-gauge part provides 
the topological characterization of the configurations 
irrespective of the detailed structure of the core. 

Assume that two gauge field configurations  $A_\mu(x)$ 
and $A^\prime_\mu(x)$ differ by such a singular gauge 
transformation $V(x)$, and denote 
the path ordered exponentials of $A_\mu$ and $A^\prime_\mu$   
around a loop $C$ by $U[C]$ and $U^\prime[C]$, respectively.  
Then $\tr\,U^\prime[C]=z\,
\tr\,U[C]$, where $z\neq1$ is a nontrivial element of the center,
whenever $V$ has obstruction $\V$ linking with the loop 
$C$; otherwise, $z=1$. Conversely, changes in the value 
of $\tr U[C]$ by elements of the center can be undone by 
singular gauge transformations on the gauge field 
configuration linking with the loop $C$. This means that 
vortex configurations are topologically characterized by elements 
of $\pi_1(SU(N)/Z(N))=Z(N)$. Thus the fluctuation 
in the value of $\tr U[C]$ by elements of $Z(N)$ expresses 
the changes in the number (mod $N$) of vortices linked with the 
loop over the set of configurations for which it is evaluated.

On the lattice , the surface $\V$ of codimension 2 is regulated 
to a coclosed set $\V$ of 
plaquettes (2-cells), i.e. a closed set of dual $(d-2)$-cells 
on the dual lattice:  a closed loop 
of dual bonds (1-cells) in $d=3$; a closed 
2-dimensional surface of dual plaquettes (2-cells) in $d=4$, 
and so on. This represents the core of a thin vortex, each 
plaquette in $\V$ carrying flux $z\in Z(N)$. 
The probability of excitation of such a thin vortex 
is suppressed by the measure at large $\beta$ with a 
cost proportional to the size of $\V$. This can 
be proven, at finite $N$, quite generally, and rigorously, 
for any reflection positive action (such as the Wilson action) 
by so-called `chessboard' estimates.

Thick vortex configurations can be constructed by
perturbing the bond variables $U_b$ in the boundary of each 
plaquette $p$ in $\V$ so as to cancel the flux $z$ on $p$, and 
distribute it over the neighboring plaquettes. Continuing this 
process by next perturbing bonds in the neighboring $p$'s 
one may distribute the flux over a larger region in the 
two directions transverse to $\V$. Beyond the thickness of 
the core, the vortex 
contribution reduces to the multivalued pure gauge. 
If $\V$ is extended enough, the vortex may be
made thick enough, so that each plaquette receives a 
correspondingly tiny portion of the original flux $z$ that 
used to be on each $p$ in $\V$. Long thick vortices may 
therefore be introduced in $\{U_b\}$ 
configurations having $\tr U_p\sim 1$ for all $p$ on 
$\Lambda$. Thus they may survive at weak coupling 
where the plaquette action becomes highly peaked around 
$\tr U_p\sim 1$. Long vortices may link with a large Wilson loop 
anywhere over the area bounded by the loop, thus potentially 
disordering the loop and leading to confining behavior. 
Thin vortices, on the other hand, necessarily incur a   
cost proportional to the size of $\V$ as noted above, 
and only short ones 
can be expected to survive at weak coupling. These can link 
then only along the perimeter of a large loop generating only 
perimeter effects.

Hybrid vortex configurations having a thick and a thin part 
are also possible. At weak coupling hybrid 
vortices survive if the thin part is short. Such hybrid vortices 
formed by long thick vortices `punctured' by a short (e.g. 
one-plaquette-long) thin part may then disorder a large Wilson 
loop in essentially the same way as thick vortices.\footnote{We 
refer to \cite{KT1} for discussion and mathematical formulation.}

\begin{figure}[h]
\begin{center}
\leavevmode
\epsfxsize=90mm 
\epsfysize=85mm
\epsfbox{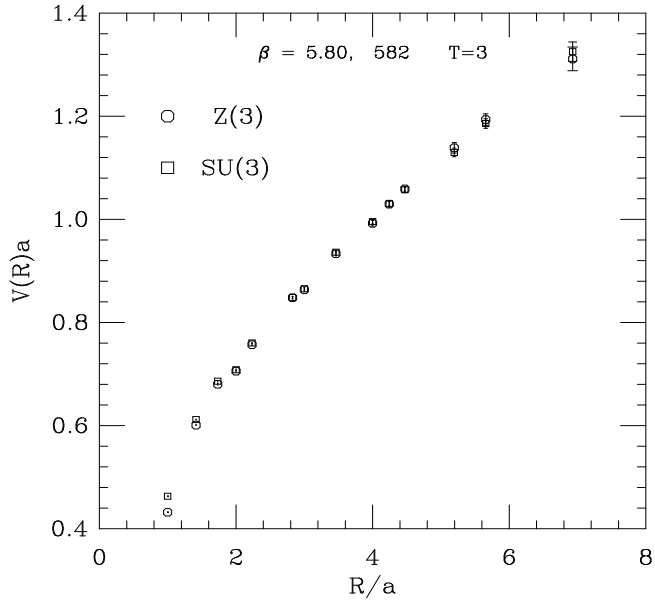}
\end{center}
\caption{The heavy quark potential at $\beta=5.8$ on a set of
582 $8^3*12$ lattices extracted at time slice T=3.}
   \label{fig:potsu3_b5.8}
\end{figure}

We separate out the $Z(N)$ part of the Wilson loop observable 
by writing $\arg (\tr U[C]) = \varphi[C]+ {2\pi \over N}n[C]$, 
where $-\pi/N < \varphi[C]\leq \pi/N$, and $n[C]=0,1,\ldots,N-1$. 
Thus, with $\eta[C]=\exp(i{2\pi \over N}n[C]) \in Z(N)$, 
\bea 
W[C] = \vev{\tr U[C]} 
          & = & \vev{|\tr U[C]|\,e^{i\varphi[C]}\,\eta[C]}\label{WL}\\
       &=& \vev{ |\tr U[C]|\,\cos(\varphi[C]+ {2\pi \over N}n[C])} 
\nonumber\\
         & =& \vev{ |\tr U[C]|\,\cos(\varphi[C])\,
               \cos({2\pi \over N}n[C])}\;, \label{WLdec}
\eea 
where the last two equalities follow from the fact that the 
expectation is real by reflection positivity, and that it is 
invariant under $n[C] \to (N-n[C])$.  We next define 
\beq 
W_{Z(N)}[C] = \vev{\cos({2\pi \over N}n[C])}\; \label{ZnWL}
\eeq 
for the expectation of the $Z(N)$ part, which, as noted above,  
gives the response to the fluctuation in the number (mod $N$) 
of vortices linking with the loop. In the following we 
compare the string tension extracted from the full Wilson loop 
$W[C]$, eq. (\ref{WL}), to the string tension extracted from 
$W_{Z(N)}[C]$, eq. (\ref{ZnWL}), for $N=3$. 
\begin{figure}[h]
\begin{center}
\leavevmode
\epsfxsize=90mm
\epsfysize=85mm
\epsfbox{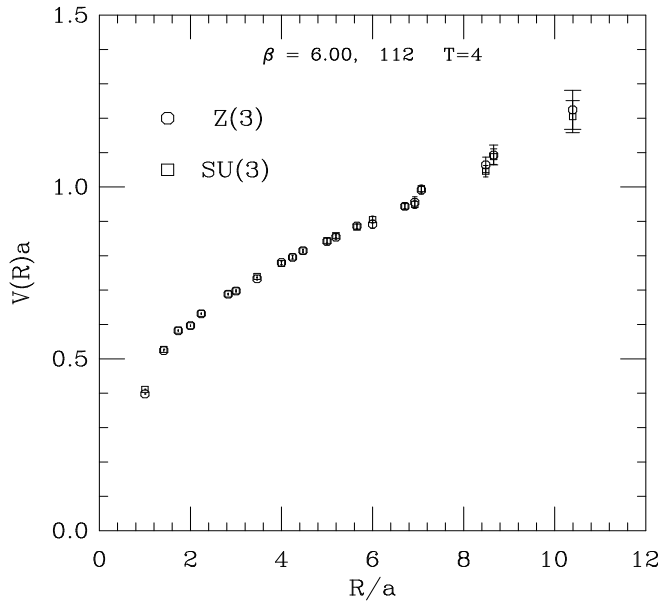}
\end{center}
\caption{The heavy quark potential at $\beta=6.0$ on a set of
112 $12^3*16$ lattices extracted at time slice T=4.}
   \label{fig:potsu3_b6.0}
\end{figure} 

It should be noted that $W_{Z(N)}[C]$, as defined by (\ref{ZnWL}), 
includes the effect of all vortices, thick and thin. As discussed, 
only thick vortices are expected to affect the string tension of 
sufficiently large loops. We will explicitly address this question 
by employing smoothing procedures that smooth out the short-distance 
fluctuations.  

\begin{figure}[h]
\begin{center}
\leavevmode
\epsfxsize=90mm
\epsfysize=85mm
\epsfbox{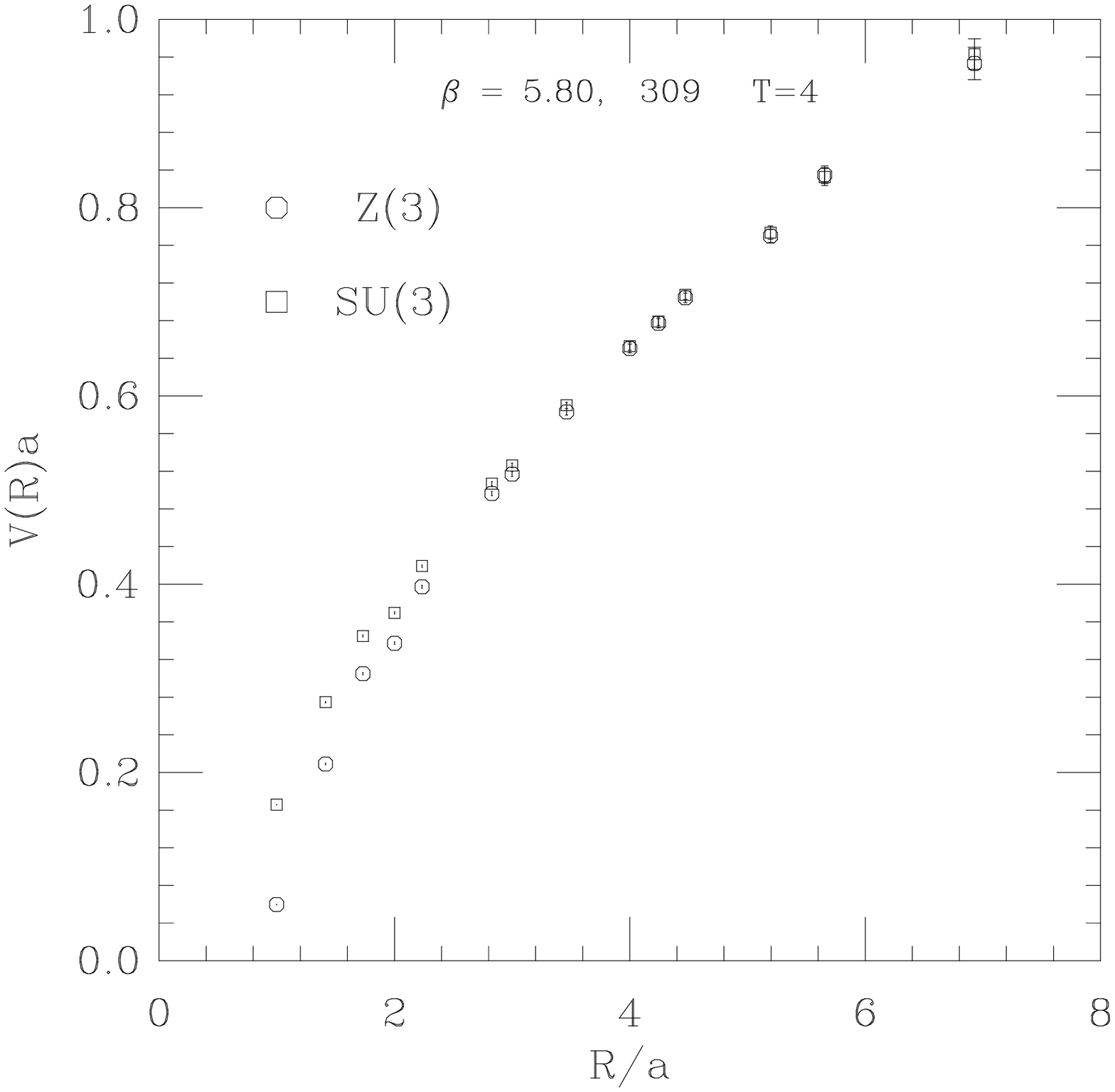}
\end{center}
\caption{The heavy quark potential at $\beta=5.8$ on a set of
582 $8^3*12$ lattices extracted at time slice T=4 on 2 times 
smoothed configurations.}
   \label{fig:potsu3_b5.8_b2}
\end{figure}
\begin{figure}[htb]
\begin{center}
\leavevmode
\epsfxsize=90mm
\epsfysize=85mm
\epsfbox{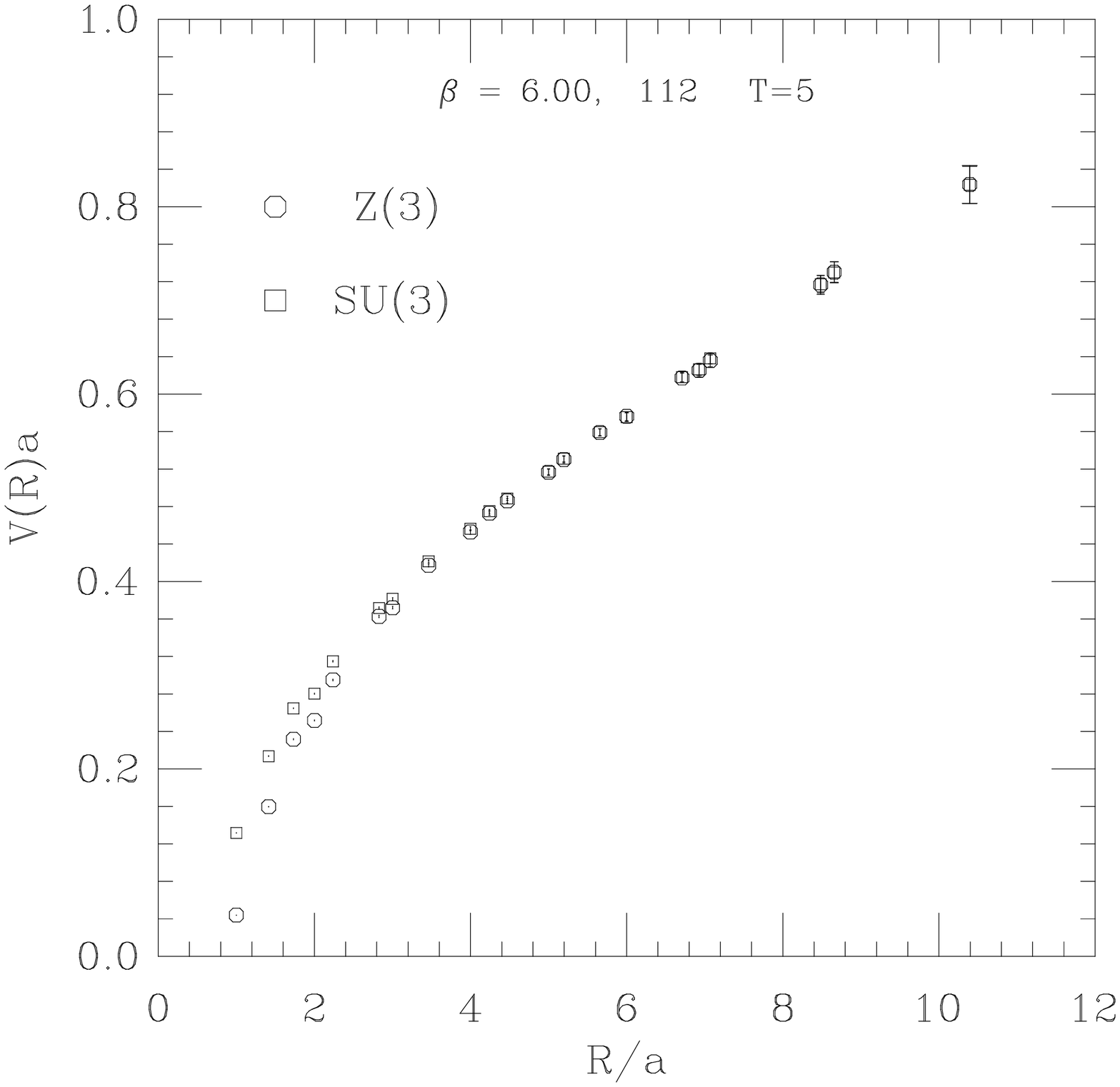}
\end{center}
\caption{The heavy quark potential at $\beta=6.0$ on a set of
112 $12^3*16$ lattices extracted at time slice T=4 from 2 times
smoothed lattices.}
   \label{fig:potsu3_b6.0_b2}
\end{figure}

\begin{figure}[htb]
\begin{center}
\leavevmode
\epsfxsize=90mm
\epsfysize=85mm
\epsfbox{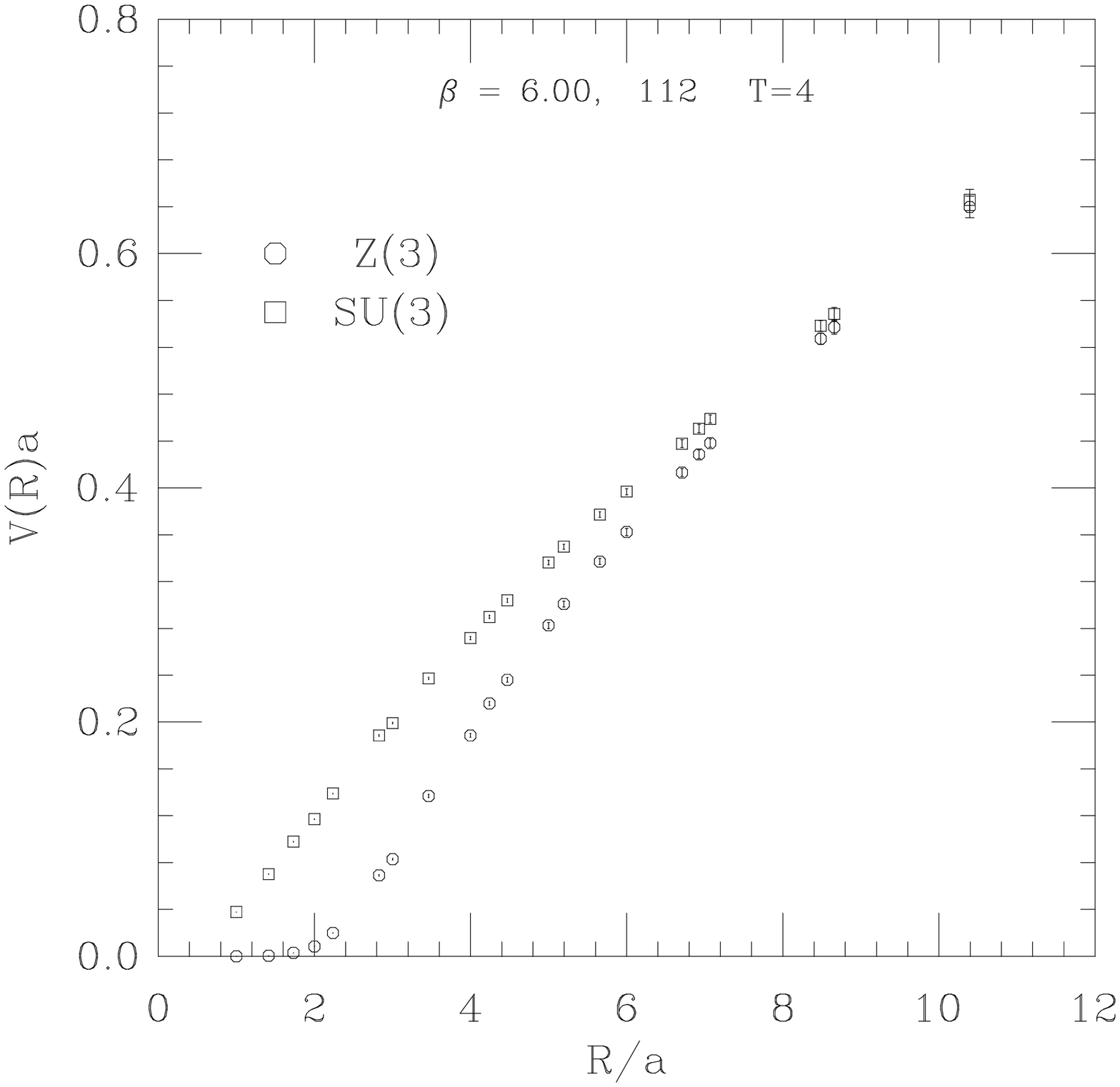}
\end{center}
\caption{The heavy quark potential at $\beta=6.0$ on a set of
112 $12^3*16$ lattices extracted at time slice T=4 from 6 times
smoothed lattices.}
   \label{fig:potsu3_b6.0_b6_t4}
\end{figure}
\begin{figure}[htb]
\begin{center}
\leavevmode
\epsfxsize=90mm
\epsfysize=85mm
\epsfbox{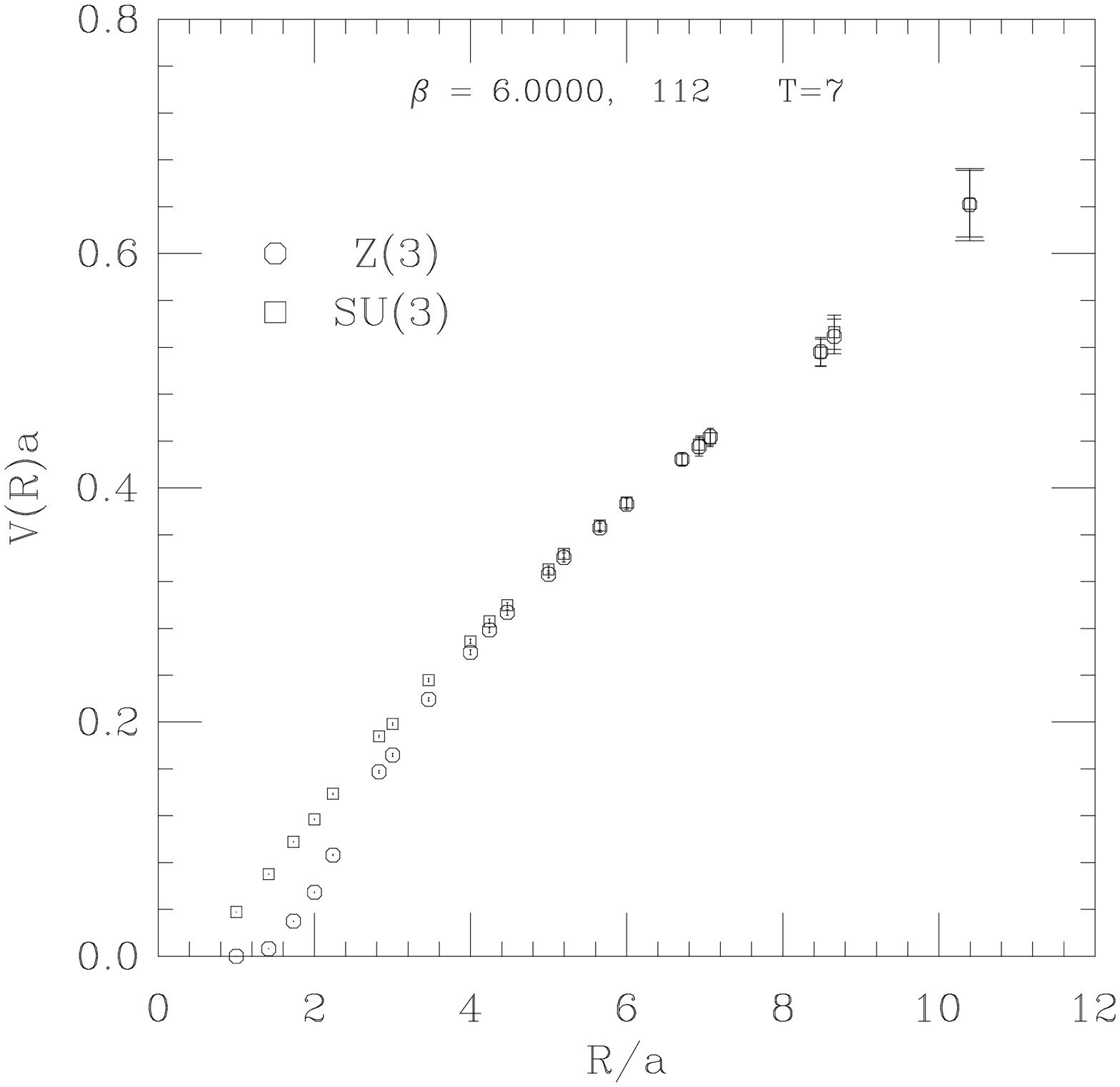}
\end{center}
\caption{Same as Fig.\ \ref{fig:potsu3_b6.0_b6_t4} except at T=7. }
   \label{fig:potsu3_b6.0_b6_t7}
\end{figure}

The heavy quark potential was extracted following \cite{Het}. 
Both on and off axis loops were computed, and, for loop $C$ of space 
extension $R$ and time extension $T$, the potential obtained from 
\beq
V(R,T) = -\ln {W(R,T+1) \over W(R,T)}\; .\label{V}
\eeq
The heavy quark potential is formally obtained by the large $T$ 
limit. Similarly, we define the potential extracted from the 
$Z(3)$ fluctuation expectation by replacing $W[C]$ in (\ref{V}) 
by $W_{Z(3)}[C]$. In the following the potentials are always 
displayed only 
after they have reached a good plateau, which, at our values of 
$\beta$, already happens at time extension of a few lattice 
spacings. We worked with the Wilson action at lattice spacings 
$a=0.15$ fm and $a=0.10$ fm for $\beta=5.8$ and $\beta=6.0$, 
respectively. This is computed from the string tension
assuming that its physical value is 440MeV.

The results are presented in Figs. \ref{fig:potsu3_b5.8} and 
\ref{fig:potsu3_b6.0}. The agreement between the potential   
extracted from the full Wilson loop and that from the $Z(3)$ 
fluctuation expectation (\ref{ZnWL}) is striking. Note that it 
includes also the short-distance regime. This is because, 
as noted above, (\ref{ZnWL}) counts {\it both} thick and thin 
vortices, and the thin ones are clearly important at short distances 
(narrow loops).\footnote{For all size 
Wilson loops thin vortices contribute significantly to the length 
piece, i.e. the constant term in the potential.}  At longer distances, 
however, only sufficiently thick vortices can be expected to 
contribute to the string tension.

To test this further we performed local smoothing on our 
configurations which removes short distance fluctuations but 
preserves the long distance physical features. If the 
string tension is really fully reproduced by the vortex 
fluctuations, the agreement seen in Figs. 
\ref{fig:potsu3_b5.8}, \ref{fig:potsu3_b6.0} should persist 
at long distances when the potential is measured  
on the smoothed configurations. 
This is in fact a very stringent test which appears to be 
routinely failed by several other recent attempts to 
isolate excitations responsible for long-distance physics. 
We used the smoothing procedure of Ref. \cite{DeGet} applied 
here to $SU(3)$.\footnote{This utilizes the type of 
smoothing, sometimes refered to as APE smearing, first 
described in \cite{Fet}.} Results for the potentials on 
twice smoothed configurations are given in Figs.   
\ref{fig:potsu3_b5.8_b2} and \ref{fig:potsu3_b6.0_b2}.     

We see that the potentials extracted from $W[C]$ and 
$W_{Z(3)}[C]$ now disagree over short distances, but then again 
merge together with no discernible difference at distances 
$R/a > 3$. This is as expected: smoothing destroys thin 
vortices but leaves vortices thicker than the smoothing 
scale unaffected.

Performing further smoothing steps extends the 
distance scale over which fluctuations are smoothed, thus 
progressively destroying vortices of larger sizes; but the 
asymptotic string tension should not be affected, since, for 
sufficiently large loops, there is a scale beyond which linked 
vortices are not affected. This is rather strikingly illustrated by 
comparing Fig \ref{fig:potsu3_b6.0_b2} to Fig. 
\ref{fig:potsu3_b6.0_b6_t4} which displays the potentials 
resulting on six times smoothed configurations.  
Alternatively, for a fixed amount of smoothing, increasing the 
time extension $T$ should enlarge the range of agreement 
of the full and $Z(3)$ potentials to include shorter distances.   
Indeed, as $T$ increases, more linked vortices can now survive the 
smoothing by being allowed to thicken in the enlarged time  
direction. This is seen in Fig. 
\ref{fig:potsu3_b6.0_b6_t7} compared to Fig. 
\ref{fig:potsu3_b6.0_b6_t4}. A smooth evolution    
extrapolating between these two figures is found as $T$ 
is increased from $4$ to $7$.

In conclusion, our numerical simulations show the 
$SU(3)$ string tension to be fully reproduced by the expectation 
of the $Z(3)$ fluctuation of the Wilson loop observable. 
The result is stable under multiple smoothings of the 
configurations removing short distance fluctuations. 
This indicates that it represents an actual long-distance 
physical feature. More generally, all our findings appear  
consistent with a physical picture of locally smooth extended 
thick vortices occuring over all long scales, and giving rise 
to the full asymptotic string tension.


\begin{thebibliography}{99} 
\bibitem{KT1} T. G. Kov\'acs, E. T. Tomboulis, Phys. Rev. {\bf D57} 
(1998) 4054; Nucl. Phys. {\bf B} (Proc. Suppl.) {\bf 63} (1998) 534; 
ibid {\bf53} (1997) 509. 
\bibitem{Det} L. Del Debbio, M. Faber, J. Greensite, \v{S}. Olejn\'{\i}k,  
Phys. Rev. {\bf D55} (1997) 2298; Nucl. Phys. {\bf B} (Proc. 
Suppl.) {\bf 63} (1998) 552;\\ 
L. Del Debbio et al, (1998) hep-lat/9801027.
\bibitem{KT2} T. G. Kov\'acs, E. T. Tomboulis, (1998) hep-lat/9806030. 
\bibitem{Het} U.M.~Heller, K.M.~Bitar, R.G.~Edwards and A.D.~Kennedy, 
Phys.\ Lett.\  {\bf B 335} (1994) 71.
\bibitem{DeGet} T. DeGrand, A. Hasenfratz, and T.G. Kov\'acs, (1997) hep-lat/9711032. 
\bibitem{Fet} M. Falcioni, M. Paciello, G. Parisi, and B. Taglienti, 
Nucl. Phys. B251[FS13] (1985) 624;\\
M. Albanese et al, Phys. Lett. B192 (1987) 163. 
\end{thebibliography}
\end{document}